\documentclass[5p]{elsarticle}

\usepackage{setspace}
\usepackage{amsmath,amssymb,amsfonts,amsthm}
\usepackage{graphicx}
\usepackage{enumerate} % advanced enumerate environment
\usepackage{colordvi} % for color text

\def\be{\begin{equation}}
\def\ee{\end{equation}}
\def\ba{\begin{eqnarray}}
\def\ea{\end{eqnarray}}

\def\dd{{\rm d}}
\def\rcr{\rho_{\rm crit}}
\def\rpl{\rho_{\rm Pl}}
\def\L{\rm L}
\def\N{\mathcal{N}}
\def\B{\rm B}

\def\pphi{p_{(\phi)}}

\def\lp{{\ell}_{\rm Pl}}
\def\R{\mathbb{R}}
\def\S{\mathbb{S}}
\def\rcr{\rho_{\mathrm{crit}}}

\def\b{$\bullet\,\,\,\, $}
\def\f{\frac}
\usepackage{enumerate}

\usepackage{colordvi}
\newcounter{mnotecount}[section]

\newcommand{\comment}[1]{}
\begin{document}

\title{Loop quantum cosmology and slow roll inflation}

\author[igc]{Abhay Ashtekar}\ead{ashtekar@gravity.psu.edu}
\author[igc]{David Sloan} \ead{sloan@gravity.psu.edu}
\address[igc]{Institute for Gravitation and the Cosmos \& Physics
 Department,
 Penn State, University Park, PA 16802-6300, USA}
\begin{abstract}

In loop quantum cosmology (LQC) the big bang is replaced by a
quantum bounce which is followed by a robust phase of
super-inflation. Rather than growing unboundedly in the past, the
Hubble parameter \emph{vanishes} at the bounce and attains a
\emph{finite universal maximum} at the end of super-inflation. These
novel features lead to an unforeseen implication: in presence
of suitable %inflationary
potentials all LQC dynamical trajectories are funneled to conditions
which virtually guarantee slow roll inflation with more than 68
e-foldings, {without any input from the pre-big bang regime}. This
is in striking contrast to certain results in general relativity,
where it is argued that the a priori probability of obtaining a slow
roll with 68 or more e-foldings is suppressed by a factor
$e^{-204}$.
%slow
%roll inflation with $\N$ e-foldings is suppressed by a factor
%$e^{-3\N}$.

\end{abstract}

%\pacs{04.60.Kz,04.60Pp,98.80Qc,03.65.Sq}

\maketitle

\section{Introduction}
\label{s1}

Inflationary models have had striking successes, especially in
providing a natural explanation of structure formation. These
successes bring to forefront an old question: Does a
sufficiently long, slow roll inflation require fine tuning of
initial conditions or does it occur generically in a given
theoretical paradigm? (See
e.g. [1 - 4]) %\cite{hw,klm,hhh,gt}).
Such a slow roll requires that initially
the inflaton must be correspondingly high-up in the potential. How
did it get there? Is it essential to invoke some rare quantum
fluctuations to account for the required initial conditions because
the a priori probability for their occurrence is low? Or, is a
sufficiently long, slow roll inflation robust in the sense that it
is realized in `almost all' dynamical trajectories of the given
theory?

To make these questions precise, one needs a stream-lined framework
to calculate probabilities of various occurrences \emph{within a
given theory}. A mathematically natural framework to carry out this
analysis was introduced over two decades ago (see e.g. [5 - 7]).
%\cite{ghs,dp,hp}).
It invokes Laplace's principle of indifference
\cite{psdl} to calculate the \emph{a priori} probabilities for
various occurrences. More precisely, the idea is to use (a flat
probability distribution $P(s)=1$ and) the canonical Liouville
measure $\dd \mu_{\L}$ on the space $\mathbb{S}$ of solutions $s$ of
the theory under consideration to calculate the \emph{relative
volumes} in $\mathbb{S}$ occupied by solutions with desired
properties \cite{ghs}. In our case, then, the a priori probability
is given by the \emph{fractional} Liouville volume occupied by the
sub-space of solutions in which a sufficiently long, slow roll
inflation occurs. Further physical input can provide a sharper
probability distribution $P(s)$ and a more reliable likelihood than
the `bare' a priori probability. However, a priori probabilities can
be directly useful if they are very low or very high. In these
cases, it would be an especially heavy burden on the fundamental
theory to come up with the physical input that significantly alters
them.

There is however a conceptual obstacle in this calculation
because of the initial singularity in general relativity, where
the matter density and curvature both diverge: there is no
clean starting point to begin one's counting of e-foldings. For
definiteness consider the standard $m^2\phi^2$ potential. If we
allow arbitrarily high energy densities at the onset of
inflation, then we have to allow initial configurations in
which the potential energy is arbitrarily large, i.e.,
initially the inflaton is arbitrarily high-up in the potential.
Then it is easy to achieve a long slow roll.
%the a priori probability of a sufficiently long slow roll would be
%high.
However, one cannot really trust general relativity at arbitrarily
high densities and curvatures. Therefore it is not clear that this
conclusion is physically reliable. Thus, because of the initial
singularity, we know we cannot trust general relativity in certain
regimes but the theory itself does not provide clear guidance to
restrict the possible initial conditions; it does not have an
in-built mechanism to determine its domain of validity.

In addition, calculation of the a priori probability can be subtle
because the total Liouville measure of the space of all solutions is
often infinite \cite{hp}. However, sometimes it is possible to
overcome this difficulty by introducing physically motivated
regularization schemes and show that the desired probability is
insensitive to the details of the scheme. Recently, this strategy
was used by Gibbons and Turok \cite{gt} to argue that the
probability of $\N$ e-folds of a slow roll, single field inflation
is suppressed by a factor of $e^{-3\N}$ in general relativity. They
concluded that, even if a cosmological model in general relativity
allows inflation, one must invoke an extremely sharp probability
distribution $P(s)$ order to explain \emph{why inflation actually
occurred}; ``the question of why and how inflation started remains a
deep mystery and a challenge for the fundamental theory.''

Loop quantum cosmology (LQC) provides a new arena to analyze this
issue because the big bang singularity is naturally resolved and
replaced by a big bounce due to quantum geometry effects [9 - 13].
%\cite{aps1,aps3,acs,apsv,kv}.
Now the question can be posed in an unambiguous way because all
solutions are regular.
%Furthermore, LQC tells us that general relativity is an
%excellent approximation once the matter density $\rho$ falls below
%$\sim\, 10^{-2}\,\rho_{\rm Pl}$. Therefore, one can start counting
%the e-foldings starting from the bounce point or, alternatively,
%from the epoch when the dynamical trajectory has reached general
%relativity regime.
Therefore one can start counting the number of e-foldings from
the bounce. The matter density operator has a \emph{finite,
universal upper-bound} \cite{acs}, whence there is an absolute
upper bound on how high the inflaton can be up the potential.
An unambiguous question now is: Can there still be sufficiently
large number of slow roll e-foldings?

The purpose of this communication is to report the main result
of a detailed analysis of these questions: In presence of
suitable potentials, every solution enjoys an inflationary
phase and the a priori probability of obtaining at least 68
e-foldings, desired from phenomenological considerations, is
extremely close to 1. Thus, the conclusion reached by Gibbons
and Turok in general relativity is reversed in LQC. Away from
the Planck regime, LQC is virtually indistinguishable from
general relativity. However, in the Planck regime, there are
huge differences and these are crucial to our analysis. In
particular, since the big bang is replaced by a non-singular
big bounce, initial conditions can be specified at the bounce
in a fully controlled fashion. There is a robust phase of
\emph{super-inflation} immediately after the bounce
\cite{mb2,ps2}. Somewhat surprisingly, it shepherds most of the
LQC solutions to phase space regions from which a long, slow
roll, expansion is almost inevitable. Although several
phenomenological consequences of the distinguishing features of
LQC have been studied (see, e.g., [16
- 19]), %\cite{nott,nott2,gre1,gre2}),
implications on slow roll inflation have not received as much
attention. To our knowledge there have been only two
investigations along these lines. The first \cite{gns} is aimed
at calculating a priori probabilities, as in this Letter, but
the bounce and super-inflation were ignored. These were
considered in \cite{jm} but systematic calculation of
probabilities, e.g. through the use of a measure, was not
carried out. In terms of the natural Liouville measure, the
solutions considered there correspond only to a \emph{very}
small region of $\mathbb{S}$.

The material is organized as follows. Section \ref{s2} recalls
the salient features of LQC that are used in our analysis.
Section \ref{s3} summarizes the technical results on the
super-inflation and inflation phases of LQC dynamics. This
discussion of dynamics in the Planck era and during the
subsequent slow roll will have applications well beyond the
main conclusions of this Letter, e.g., in the analysis of how
perturbations evolve during the bounce. Section \ref{s4} uses
the Liouville measure to show that, in presence of a suitable
potential, the a priori probability of inflation is \emph{very}
close to 1. Section \ref{s5} compares and contrasts our methods
and results with those in the literature.

\section{Loop quantum cosmology: Relevant results}
\label{s2}

In the LQC treatment of simple cosmological models, the big bang and
big crunch singularities are naturally resolved \cite{mb1}. The
origin of this resolution lies in the \emph{quantum geometry
effects} that are at the heart of loop quantum gravity [23 - 25]. %\cite{alrev,crbook,ttbook}.
Exotic matter is not needed; indeed matter fields can satisfy all
the standard energy conditions. Detailed analysis has been carried
out in a variety of models: the k=0, $\pm$1\,
Friedmann-Lema\^itre-Robertson-Walker (FLRW) space-times with or
without a cosmological constant \cite{aps3,apsv,kv,bp}; Bianchi
models [27 - 29] %\cite{awe2,awe3,we}
which admit anisotropies and gravitational waves; and Gowdy models
\cite{gowdy} which admit inhomogeneities, and therefore an infinite
number of degrees of freedom. The FLRW models have been studied most
extensively, using both analytical and numerical methods to solve
the exact quantum equations of LQC [10 - 13]. %\cite{aps3,acs,apsv,kv}.
In these models, the big bang and the big-crunch are replaced by a
quantum bounce, which is followed by a robust phase of
super-inflation. Interestingly, full quantum dynamics, including the
bounce, is well-approximated by certain effective equations. (For a
recent review, see \cite{aa-badhonef}.)

In this Letter we restrict ourselves to the phenomenologically
more interesting case of the k=0 FLRW model (although the
method is applicable also to the k=-1 case). The matter source
will be a scalar field with positive kinetic energy and a
suitable potential. Since all the prior discussion of
probabilities is based on general relativity, to facilitate
comparison we use effective equations rather than the full
quantum theory. Finally, we will use the natural Planck units
c=$\hbar$=G=1 (rather than $8\pi$G=1, often employed in
cosmology). The fundamental time unit, $\sqrt{G\hbar/c^5}$,
will be referred to as a \emph{Planck second}.

In LQC, spatial geometry is encoded in the volume of a fixed,
fiducial cell, rather than the scale factor $a$;\, $v = ({\rm
const})\times a^3$. The conjugate momentum is denoted by $b$. On
solutions to Einstein's equations, $b= \gamma H$ \cite{acs}. (Here
$H = \dot{a}/a$ is the standard Hubble parameter and $\gamma$ is the
Barbero-Immirzi parameter of LQC whose value, $\gamma \approx 0.24$,
is fixed by the black hole entropy calculation.) However, LQC
modifies Einstein dynamics and on solutions to the LQC effective
equations we have
\be \label{H} H \,=\, \f{1}{2\gamma\lambda}\, \sin 2\lambda b
\,\,\approx\,\, \f{0.93}{\lp}\, \sin 2\lambda b \ee
where $\lambda^2 \approx 5.2 \lp^2$ is the `area-gap', the smallest
non-zero eigenvalue of the area operator. In LQC,  $b$ ranges over
$(0, \pi/\lambda)$ and general relativity is recovered in the limit
$\lambda \rightarrow 0$. Quantum geometry effects modify the
geometric, left side of Einstein's equations. In particular, the
Friedmann equation becomes
\be \label{lqc-fe2} \f{\sin^2 \lambda b}{\gamma^2\lambda^2}\,\,
=\,\, \f{8\pi}{3}\, \rho \,\,\equiv\,\, \f{8\pi}{3}\,
\big(\f{{\dot\phi}^2}{2} + V(\phi) \big)\, .\ee
To compare with the standard Friedmann equation $H^2 =
(8\pi/3)\,\rho$, it is often convenient to use (\ref{H}) to write
(\ref{lqc-fe2}) as
\be \label{lqc-fe} \f{1}{9}\,(\f{\dot{v}}{v})^2\, \equiv H^2 =
\f{8\pi}{3} \,\,\rho\, \big(1 - \f{\rho}{\rcr}\big) \,  \ee
where $\rcr = \sqrt{3}/32\pi^2 \gamma^3 \approx 0.41 \rpl$. By
inspection it is clear from Eqs (\ref{H}) - (\ref{lqc-fe}) that
away from the Planck regime ---i.e., when $\lambda b \ll 1$,
or, $\rho \ll \rcr$--- we recover classical general relativity.
However, modifications in the Planck regime are drastic. The
main
features of this new physics can be summarized as follows.\\

\b In general relativity the Friedmann equation implies that if
the matter density is positive, $\dot{a}$ cannot vanish.
Therefore every solution represents \emph{either} a contracting
universe \emph{or} an expanding one. By contrast, the LQC
modified Friedmann equation (\ref{lqc-fe}) implies that
$\dot{v}$ vanishes at $\rho=\rcr$. This is a quantum bounce. To
its past, the solution represents a contracting universe with
$\dot{v} <0$ and to its future, an expanding one with $\dot{v}
>0$.

\b As is customary in the literature on probabilities, let us
ignore the exceptional de Sitter solutions with never-ending
inflation. On all other solutions $b$ decreases monotonically
from $b = \pi/\lambda$ to $0$. Eqs (\ref{lqc-fe2}) and
(\ref{lqc-fe}) imply that $b=\pi/2\lambda$ at the bounce. Thus,
each solution undergoes precisely one bounce. The Hubble
parameter $H =\dot{v}/3v$ \emph{vanishes} at the bounce and
Eq.(\ref{H}) implies that it is bounded on the full solution
space $\mathbb{S}$;\, $|H| \lesssim 0.93/\lp$. By contrast, in
general relativity, $H$ is large in the entire Planck regime
and diverges at the singularity.

\b If the potential $V(\phi)$ is bounded below, say $V \ge V_o$,
then it follows from (\ref{lqc-fe2}) that ${\dot\phi}^2$ is bounded
by $2\rcr - 2V_o$. If $V$ grows unboundedly for large $|\phi|$, then
$|\phi|$ is also bounded. For example, for $V= m^2\phi^2/2$, we have
$m|\phi|_{\rm max} = 0.90$.

\b When the potential is bounded below, $|\dot{H}|$ is bounded above
by $10.29/\lp^2$. The Ricci scalar ---the only non-trivial curvature
scalar in these models--- is bounded above by $31/\lp^2$. Thus,
physical quantities which diverge at the big bang of general
relativity cannot exceed certain finite, maximum values in LQC. One
can also show that if $v \not=0$ initially, it cannot vanish in
finite proper time along any solution. Thus, \emph{the LQC solutions
are everywhere regular irrespective of whether one focuses on matter
density, curvature or the scale factor.} \\

Next, the full set of space-time equations of motion can be written
in terms of $v(t),\phi(t)$. These variables are subject to the
constraint (\ref{lqc-fe}) and evolve via:
\ba \label{dyn} &&\ddot{v} = \f{24\pi v}{\rcr}\,\big[(\rho
-V(\phi))^2 + V(\phi) (\rcr-V(\phi))\big]\nonumber\\
&&\ddot\phi + \f{\dot{v}}{v}\, \dot\phi + V_{,\phi} =0\, . \ea
Our task is to obtain the Liouville measure on the space $\S$ of
solutions to these equations.

For this, we first construct the phase space $\Gamma$. It consists
of quadruplets $(v,b;\, \phi, \pphi)$, with $\lambda b \in [0,
\pi/2]$. The Liouville measure on $\Gamma$ is simply $\dd\mu_{\L} =
{\dd} {v}\, {\dd} {b}\, \dd\phi\, {\dd}\pphi$. The LQC Friedmann
equation implies that these variables must lie on a constraint
surface $\bar\Gamma$ defined by
\be \label{hc} \f{3\pi}{2\lambda^2}\, \sin^2\lambda b \,
=\,\f{\pphi^2}{2v^2} +4\pi^2\gamma^2 V(\phi)\, . \ee
They evolve via
\ba\label{evo}
 \dot{v} &=& \f{3v}{2\gamma}\, \f{\sin 2\lambda b}{\lambda}, \quad
 \dot{b} = - \f{\pphi^2}{\pi \gamma v^2},\nonumber\\
 \dot\phi &=& \f{\pphi}{2\pi\gamma v},  \quad\quad\quad
 {\dot p}_{(\phi)} = -2\pi\gamma |v|\, V_{,\phi}\, . \ea
As is well-known, the space of solutions $\S$ is naturally
isomorphic to a gauge fixed surface, i.e., a 2-dimensional surface
$\hat\Gamma$ of $\bar\Gamma$ which is intersected by each dynamical
trajectory once and only once. Since $b$ is monotonic in each
solution, an obvious strategy is to choose for $\hat\Gamma$ a
2-dimensional surface $b= b_o$ (a fixed constant) within
$\bar\Gamma$. Symplectic geometry considerations unambiguously equip
$\hat\Gamma$ ---and hence the solution space $\S$--- with an induced
Liouville measure $\dd\hat\mu_{\L}$. Since the dynamical flow
preserves the Liouville measure, $\dd\hat\mu_{\L}$ on $\S$ is
\emph{independent of the choice of} $b_o$. The most natural choice
in LQC is to set $b_o = \pi/2\lambda$ so that $\bar\Gamma$ is just
the `bounce surface'. We will make this choice because it also turns
out to be convenient for calculations.

Then $\hat\Gamma$ is naturally coordinatized by
$(\phi_{\B},v_{\B})$, the scalar field and the volume at the bounce.
Since $b= \pi/2\lambda$, the constraint (\ref{hc}) determines
$\pphi$ (or, equivalently, $\dot\phi$) up to sign which, without
loss of generality, will be taken to be non-negative. The induced
measure on $\S$ can be written explicitly as:
\be \dd\hat\mu_{\L} =   \f{\sqrt{3\pi}}{\lambda}\,\, \big[1-
F_{\B}\big]^{\f{1}{2}}\,\, {\dd} \phi_{\B}\, {\dd} {v}_{\B} \ee
where $F_{\B} = V(\phi_{\B})/\rcr$ is the fraction of the total
density that is in the potential energy at the bounce. The total
Liouville volume of $\bar\Gamma \equiv \S$ is infinite because,
although $\phi_{\B}$ is bounded for suitable potentials such as
$m^2\phi^2$, $v_{\B}$ is not. However, this non-compact direction
represents gauge on the space of solutions $\S$: If $(\phi(t),
v(t))$ is a solution to (\ref{lqc-fe2}) and (\ref{dyn}), so is
$(\phi(t),\alpha v(t))$ and this rescaling by a constant $\alpha$
simply corresponds to a rescaling of spatial coordinates (or of the
fiducial cell) under which physics does not change. Therefore, as
discussed in section \ref{s4}, there is a natural prescription to
calculate fractional volumes of physically relevant sub-regions of
$\hat\Gamma$ by factoring out the gauge orbits.

\section{Super-inflation and inflation}
\label{s3}

\begin{center}
% use packages: array
\newcommand{\mc}[3]{\multicolumn{#1}{#2}{#3}}

\begin{table*}

\hskip1.2cm{
\begin{tabular}{|l | c | r | r | r | r | r |}
\hline
{$F_B=V(\phi_{\B})/\rcr$} &  {Sign[$\phi_{\B}$]} &  {$t$} &  {$\phi$} &  {$\dot{\phi}$} &  {$H$} &  {$\dot{H}$} \\  \hline
{$0$} &  {+/-} &  {$1.6*10^{6}$} &  {$2.3$} &  {$-9.7*10^{-8}$} &  {$2.8*10^{-6}$} & {$-1.2*10^{-13}$} \\  \hline
{$4.4*10^{-13}$} &  {+} &  {$1.2*10^{6}$} &  {$3.2$} &  {$-9.7*10^{-8}$} &  {$4.0*10^{-6}$} &  {$-1.2*10^{-13}$} \\
{} &  {-} &  {$2.1*10^{6}$} &  {$1.3$} &  {$9.6*10^{-8}$} &  {$1.6*10^{-6}$} &  {$-1.2*10^{-13}$} \\ \hline
{$1*10^{-4}$} &  {+} &  {$7.6*10^{2}$} &  {$1.5*10^4$} &  {$-9.8*10^{-8}$} &  {$1.9*10^{-2}$} &  {$-1.2*10^{-13}$} \\
{} &  {-} &  {$6.6*10^{2}$} &  {$-1.5*10^4$} &  {$9.8*10^{-8}$} &  {$1.9*10^{-2}$} &  {$-1.2*10^{-13}$} \\  \hline
{$0.5$} &  {+} &  {$1.6*10^{1}$} &  {$1.1*10^6$} & {$-1.4*10^{-7}$} &  {$9.3*10^{-1}$} &  {$1.4*10^{-19}$} \\
{} &  {-} &  {$1.5*10^{1}$} &  {$-1.1*10^6$} &  {$1.4*10^{-7}$} &  {$9.3*10^{-1}$} &  {$-1.4*10^{-19}$} \\  \hline
{$0.8$} &  {+} &  {$2.0*10^{1}$} &  {$1.3*10^6$} &  {$-2.2*10^{-7}$} &  {$7.4*10^{-1}$} &  {$3.6*10^{-13}$} \\
{} &  {-} &  {$1.8*10^{1}$} &  {$-1.3*10^6$} &  {$2.2*10^{-7}$} &  {$7.4*10^{-1}$} &  {$3.6*10^{-13}$} \\ \hline
\end{tabular}
}
\caption{Values of the proper time, the Hubble parameter, the scalar
field and their time derivatives at onset of slow roll (where
$\ddot{\phi}=0$). $F_{\B} = V(\phi_{\B})/\rcr$ is the ratio of the
potential energy density to the total energy density at the bounce.
If the value $\phi_{\B}$ of the scalar field is positive, the
inflaton rises up the potential after the bounce while if
$\phi_{\B}$ is negative it descends down the potential (because
$\dot\phi_{\B}$ is assumed to be positive). For $\phi_{\B} >0$,
there are 68 e-foldings if $F_{\B} = 4.4 \times 10^{-13}$. The
bounce is taken to occur at $t=0$.}
\end{table*}
\end{center}
\vskip-1cm

For our purposes it suffices to focus just on the \emph{post bounce}
part of solutions; explicit information from the pre-bounce part is
not needed anywhere in the analysis. As explained in section
\ref{s1}, the key question is: What is the fractional Liouville
volume in $\S$ occupied by solutions that exhibit a sufficiently
long inflation? To answer it in detail, as is common in literature
(see, e.g. \cite{klm,gt}), we will use $V(\phi) = (1/2) m^2\phi^2$.
Then, as we already noted, (\ref{hc}) implies that $m\phi_{\B} \in
[-0.90, 0.90]$. For definiteness, we will use the phenomenological
value \cite{phenom}, $m = 6\times 10^{-7} {\rm M}_{\rm Pl}$ (recall
that we have set G=1 rather than $8\pi$G=1). However, as explained
in section \ref{s5}, the main results are robust even if $m$ were to
change by a couple of orders of magnitude.

The idea is to allow all possible initial conditions at the bounce
and construct dynamical trajectories by solving (\ref{evo}).  The
problem can be divided into three parts using the value of the
fraction $F_{\B}$ at the bounce. In each part, one can introduce
suitable approximations to analyze dynamics. Because the evolution
equations (\ref{dyn}) are invariant under $\phi \rightarrow -\phi,\,
\dot\phi \rightarrow -\dot\phi,\, v\rightarrow v, \dot{v}
\rightarrow \dot{v}$, it suffices to restrict ourselves to initial
data with $\dot\phi_{\B} \ge 0$ at the bounce, allowing $\phi_{\B}$
to take both positive and negative values. Let us begin with the
part $\S^+$ of solutions on which $\phi_{\B}$ is non-negative.
%at the bounce surface %$\hat\Gamma$.
Then the main results can be summarized as follows. (See also Table 1.)

(i) $F_{\B} < 10^{-4}$: \emph{Extreme kinetic energy domination
at the bounce.} At the bounce the Hubble parameter $H$
vanishes. However, there is a short phase of super-inflation
lasting a fraction of a Planck second during which $H$
increases very rapidly to its maximum value $H_{\rm max} =
0.93$. At this point $\dot{H}$ vanishes and then $H$ starts
decreasing and continues to decrease during the rest of the
evolution. Since $\dot\phi>0$, the inflaton climbs up the
potential during super-inflation and continues to do so after
super-inflation ends, till it reaches a turn-around point where
$\dot\phi=0$. Then it starts descending. Very soon after that,
$\ddot\phi$ vanishes. This is the onset of slow roll inflation:
during this phase $\dot{H}/H^2$ is in the range $1.6\times
10^{-2} - 3.3\times 10^{-10}$ so the slow roll conditions are
met. The time required to reach this onset starting from the
bounce is in the range of $10^6 - 10^2$ ${\rm s}_{\rm pl}$
where ${\rm s}_{\rm Pl}$ denotes Planck seconds. The number of
e-foldings during this slow roll is given approximately by
\be \label{N} \N \approx 2\pi\, \big(1-\f{\phi_o^2}{\phi_{\rm
max}^2}\big)\, \phi_o^2\, \ln \phi_o \ee
where $\phi_o$ is the value of the scalar field at the \emph{onset}
of inflation and $\phi_{\rm max} = 1.5 \times 10^{6}$. Now, $\phi_o$
increases monotonically with $\phi_{\rm B}$ (and is always larger
than $\phi_{\B}$). For $\phi_{\B} = 0.99$, we have $\phi_o = 3.24$
and $\N = 68$. Thus, \emph{for a kinetic energy dominated bounce,
there is a slow roll inflation with over $68$ e-foldings for all}
$\phi_{\B}>1$, i.e., if $F_{\B} > 4.4\times 10^{-13}$.

(ii) $10^{-4} < F_{\B} < 0.5$: \emph{Kinetic energy domination
at the bounce.} The LQC departures from general relativity are
now increasingly significant. The super-inflation era is
similar to case (i). However, now the value of $\phi_{\B}$ is
higher and that of $\dot\phi_{\B}$ lower while, as before, $H$
is very high at the end of super-inflation. Therefore, the
coefficient of friction, $H/m^2$, is large and one arrives at
the slow roll conditions within 10-100 $s_{\rm Pl}$ after the
bounce. Consequently, now the change $(\phi_o-\phi_{\B})$ is
negligible, a key feature not shared by regime (i). At the
onset of slow roll inflation, the Hubble parameter is now given
to an excellent approximation by
\be \label{Ho} H_o \approx \big[\, \f{8\pi}{3}\, \rcr\, F_{\B}
(1-F_{\B})\,\big]^{1/2}\, \approx \,  1.9 \, \big[\,F_{\B}
(1-F_{\B})\,\big]^{1/2}\, \ee
and decreases \emph{very} slowly with $\dot{H}/H^2 < 3.5 \times
10^{-10}$. Thus, the Hubble parameter is essentially frozen to the
value (\ref{Ho}) and the slow roll condition is met even more
easily. This value of $H$ is very high, in the range $1.9 \times
10^{-2}\,\, {\rm s}_{\rm Pl}^{-1}$ to $9.3 \times 10^{-1}\,\, {\rm
s}_{\rm Pl}^{-1}$. The Hubble freezing is an LQC phenomenon: It
relies on the fact that $H$ acquires its largest value $H_{\rm max}
= 0.93\, {\rm s}_{\rm Pl}^{-1}$ at the end of super-inflation (and,
in the case under consideration, $\dot\phi_{\B}$ is not large enough
to decrease $H$ more than two orders of magnitude). Eq. (\ref{N})
implies that throughout this range of $F_{\B}$ there are more than
68 e-foldings.

(iii) $0.5 < F_{\rm B} < 1$: \emph{Potential energy domination at
the bounce.} Now the LQC effects dominate. Again, because $\dot\phi
>0$, the inflaton climbs up the potential but now the turn around
($\dot\phi =0$) %and the onset of slow roll ($\ddot\phi =0)$
occurs \emph{during super-inflation.} The change $(\phi_o
-\phi_{\B})$ is even more negligible because the kinetic energy at
the bounce is lower than that in case (ii). The Hubble parameter
again freezes at the onset of inflation to the value given in
(\ref{Ho}). The slow roll conditions are easily met as $\dot{H}/H^2$
is less than $1 \times 10^{-11}$ when $\ddot\phi =0$ (or very soon
thereafter). A difference from the slow roll inflation of (i) and
(ii) above is that $H$ continues to grow during the slow roll
because we are in the super-inflation phase. There are many more
than 68 e-foldings already in the super inflation phase. The
inflaton exits the super-inflation phase with $H$ at its maximum
value, $H_{\rm max} =0.93$ and little kinetic energy. Therefore, the
friction term is large and the inflaton enters a long slow roll
inflationary phase. There are many more than 68 e-foldings also in
this phase.

Finally, let us consider the part $\S^-$ of the solution space on
which $\phi_B <0$. The main difference now is that the inflaton
starts rolling down the potential immediately after the bounce. As
before, in case (ii) the Hubble freezing occurs soon after the end
of super-inflation and in (iii) during super-inflation. The value of
$H_o$ is again given by (\ref{Ho}). In case (i), differences can
arise from the part $\S^+$ of the solution space because now the
kinetic energy is very large at the bounce point so the inflaton can
transit from a negative to a positive value before the onset of
inflation. But after the onset, the situation is the same as in
(ii). In this case, there are more than 68 e-foldings if $F_{\B} >
1.4\times 10^{-11}$ or $\phi_{\B} \not\in [-5.7, 0]$.

These general features of LQC dynamics emerge from analytical
calculations based on approximations that are tailored to the
three cases considered above. They were confirmed by detailed
numerical simulations performed in MATLAB using a Runge Kutta
(4,5) algorithm (ode45) to solve the set of coupled ODEs. Both
relative and absolute tolerances were set at $3\times 10^{-14}$
and the preservation of the Hamiltonian constraint (\ref{hc})
to this order was verified on each solution. To ensure
numerical accuracy, the natural logarithm of volume was treated
as fundamental in the simulations. As noted above, the
Barbero-Immirzi parameter was set at 0.24 and inflaton mass
$6\times 10^{-7}$ (in units c=$\hbar$=G=1). A large number of
simulations were performed. Table 1 summarizes a few
illustrative results.

\section{Measure and Probabilities}
\label{s4}

As explained in section \ref{s2}, the space $\S$ of solutions can be
coordinatized by pairs $(\phi_{\B}, v_{\B})$. However, physics does
not change under $(\phi_{\B}, v_{\B}) \rightarrow (\phi_{\B}, \alpha
v_{\B})$, where $\alpha$ is a constant. In particular, the number of
slow-roll e-foldings is insensitive to this rescaling of $v_{\B}$.
Therefore, only those regions $\mathcal{R}$ in $\mathbb{S}$ that
contain complete gauge orbits are physically relevant. These are of
the type $\mathcal{R} = I\times \R^+$ where $I$ is a closed interval
in $[-\phi_{\rm max},\, \phi_{\rm max}]$ and $\R^+$ denotes the
$v_{\B}$ axis. To calculate fractional volumes $P_{\mathcal{R}}$ of
such regions it is natural to factor out by the `volume of the gauge
orbits'. This suggests an obvious strategy, commonly used in the
physics literature:
\ba \label{prob} P_{\mathcal{R}}&=& \lim_{v_0 \rightarrow 0}\,\,\,
\f{\hbox{\rm Liouville Volume of}\,\, [I \times I_{v_0}]}{\hbox{\rm
Liouville Volume of}\,\, [I_{\rm total}\times I_{v_0}]}\nonumber\\
\,&=&\,\f{\int_{I}{\dd} \phi_{\B}\,
[1-F_{\B}]^{\f{1}{2}}}{\int_{-\phi{\rm max}}^{\phi{\rm max}} \dd
\phi_{\B}\,[1-F_{\B}]^{\f{1}{2}}} \ea
where we have set $I_{v_0} = [v_0, 1/v_0]$ (with $v_0>0$). This
physical idea can be mathematically justified by the `group
averaging technique' \cite{ga} to obtain a physical measure on $\S$
by averaging $\dd\hat\mu_{\L}$ over the orbit of the `gauge group.'

Let us now apply this strategy to calculate the probability that,
prior to re-heating, there are at least 68 slow roll e-foldings in
LQC. Since $F_{\B}$ ranges over $[0,1]$ and there are requisite
number of e-foldings if $F_{\B} > 1.4 \times 10^{-11}$, it follows
from (\ref{prob}) that the required probability is greater than
0.99999.  Moreover, numerical simulations show that even when
$F_{\B} \le 1.4 \times 10^{-11}$ there are at least 6.1 e-foldings
in LQC. Thus the probability of obtaining at least 6.1 e-foldings is
1. By contrast, the Gibbons and Turok result implies that in general
relativity even this probability is suppressed by a factor of
$e^{-18.3} \approx 1.1\times 10^{-8}$ \cite{gt}.

\section{Discussion}
\label{s5}

In this paper we reported the results of a systematic analysis of
LQC dynamics in the context on inflation. In LQC, all solutions of
the quantum as well as effective equations are regular and, for the
FLRW models under consideration, effective equations provide an
excellent approximation to the full quantum dynamics. To facilitate
comparison with the earlier work in general relativity, we focused
on effective equations. Every solution of these equations is
determined by its initial data at the bounce. We divided the space
of these initial data into three classes and used approximation
schemes to extract the behavior of the dynamical trajectories they
lead to. These analytical results were then confirmed by detailed
and high precision numerical calculations where both relative and
absolute tolerances were set at $3\times 10^{-14}$. By examining
\emph{all} the dynamical trajectories (not just `generic ones') we
were able to conclude that for the $m^2\phi^2$ potential with $m$
chosen to satisfy phenomenological constraints, the a priori
probability of obtaining at least 68 e-foldings is greater than
0.99999. By contrast, the Gibbons and Turok \cite{gt} argument says
that, in general relativity, this a priori probability is suppressed
by a factor $e^{-204} \sim 2.5\times 10^{-89}$!

Thus, the situation in LQC is dramatically different from that
implied by the Gibbons-Turok analysis in general relativity. Note
that we used the same potential as in the detailed calculations of
Gibbons and Turok \cite{gt}, as well as Kofman, Linde and Mukhanov
\cite{klm}. Authors of \cite{klm} have argued that a sufficiently
long, slow roll inflation will occur generically within general
relativity. Thus, the thrust of their conclusion is opposite to that
of \cite{gt}. However, they used a measure which is not preserved
under dynamics and requires an additional structure. Therefore there
has been some debate \cite{hw,klm,gt} about the appropriateness of
the procedure employed to arrive at their conclusion. In LQC one
does not have to take a stand on this issue: As in \cite{gt} we use
the natural Liouville measure which is preserved by dynamics and yet
the conclusions of \cite{gt} are reversed. Finally, the procedure we
used to handle the fact that the total Liouville volume of
$\mathbb{S}$ is infinite is physically and mathematically well
motivated and it also constituted the basis of the regularization
scheme used in \cite{gt}.

Our detailed analysis made a crucial use of the salient
differences between LQC and general relativity in the Planck
regime. Since LQC has its basis in LQG, a candidate fundamental
theory of quantum gravity, it has precise predictions in the
Planck regime of the simple cosmological models that have been
traditionally used \cite{hw,klm,hhh,gt,ghs,dp,hp} in
probability considerations. Consequently, we do not have to
worry about setting judicious initial conditions at the
singular big bang. The bounce is regular and we considered
\emph{all possible} initial data there. The LQC dynamics are
such that if $F_{\B}$, the fraction of total energy that is in
the potential at the bounce, satisfies $F_{\B} > 1.4\times
10^{-11}$ a slow roll with 68 e-foldings is inevitable. Thus,
in LQC a sufficiently long slow roll inflation may not result
\emph{only if} $F_{\B} < 1.4\times 10^{-11}$. Since by
definition $F_{\B} \in [0,1]$ for all initial conditions,
(\ref{prob}) implies that the probability of a sufficiently
long slow roll inflation is extremely close to 1.

These main results are quite robust. For example, we could change
the value of the mass used in the main calculations. The probability
of obtaining at least 68 e-foldings in fact grows slightly if $m$ is
decreased (so long as it is non-zero). What if we increase the mass?
To check robustness, let us be generous with phenomenological
constraints and increase it by \emph{two orders of magnitude},\,
i.e., require only $m < 6\times 10^{-5} M_{\rm Pl}$. Even then the a
priori probability of \emph{not} obtaining at least 68 e-foldings is
\emph{less than} $2.7 \times 10^{-4}$. Thus, we do not have to fine
tune the mass. The situation is similar with respect to adding a
quartic term to the potential with a phenomenologically permissible
coupling constant. Finally LQC provides a neat separation between
the regime in which the quantum geometry effects dominate and the
regime in which general relativity serves as an excellent
approximation. Therefore it is possible to separate the two types of
effects. These issues will be discussed in the detailed paper.

To conclude, we emphasize that we have discussed prediction of LQC
\emph{only} in presence of a scalar field with suitable potentials.
If there is no potential at all, there is still a period of
accelerated expansion due to super inflation but, unfortunately, it
does not yield a sufficient number of e-foldings. So the issue of
the origin of the required potential ---and of the inflaton
itself--- still remains. Although there have been some tantalizing
suggestions \cite{ty} that promoting the Barbero-Immirzi parameter
$\gamma$ to a field could provide a natural avenue to address these
issues, these ideas have not been analyzed in sufficient detail.

\textbf{Acknowledgments:} We would like to thank William Nelson,
Parampreet Singh and Neil Turok for discussions. This work was
supported in part by the NSF grant\\ PHY0854743 and the Eberly and
Frymoyer research funds of Penn State.

\vfill\break


\begin{thebibliography}{99}

%\bibitem{rp1}

\bibitem{hw}S. Hollands and R.M. Wald, Gen. Rel. Grav. \textbf{34},
    2043 (2002); \texttt{arXiv:hep-th/021000}

\bibitem{klm} L.A. Kofman, A. Linde and V.F. Mukhanov, J. High
    Energy Phys. \textbf{10}, 057 (2002)

\bibitem{hhh} J.B. Hartle, S.W. Hawking and T. Hertog, Phys. Rev.
    Lett. \textbf{100}, 2011301 (2007)

\bibitem{gt} G.W. Gibbons and N. Turok, Phys. Rev. D \textbf{77},
    063516 (2008)

\bibitem{ghs} G. W. Gibbons, S.W. Hawking and J. Stewart, Nucl.
    Phys. \textbf{B281}, 736 (1987)

\bibitem{dp} D.N. Page, Phys. Rev. D. \textbf{36}, 1607 (1987)

\bibitem{hp} S.~W.~Hawking and D.~N.~Page, Nucl. Phys.
    \textbf{B298}, 789 (1988)

\bibitem{psdl} P.S. de Laplace, \emph{Th\'eorie analytique des
    probabilit\'es} (Courcier, Paris, 1812); \,\,
\emph{A philosophical essay on probabilities}, translated by A. I.
Dale, (Springer, New York, 1995)

\bibitem{aps1} A.~Ashtekar, T.~Pawlowski and P.~Singh, Phys. Rev.
    Lett. \textbf{96}, 141301 (2006)

%\bibitem{aps2} A.~Ashtekar, T.~Pawlowski and P.~Singh, Phys. Rev.
%    {\bf D73}, 124038 (2006)

\bibitem{aps3} A.~Ashtekar, T.~Pawlowski and P.~Singh, Phys. Rev.
    {\bf D74}, 084003 (2006)

\bibitem{acs}A.~Ashtekar, A.~Corichi and P.~Singh, Phys. Rev.
    \textbf{D77}, 024046 (2008)

\bibitem{apsv} A.~Ashtekar, T.~Pawlowski, P.~Singh and
    K.~Vandersloot,  Phys. Rev. \textbf{D75}, 0240035 (2006);\\
L.~Szulc, W.~Kaminski, J.~Lewandowski, Class.\ Quant.\ Grav.\
\textbf{24}, 2621-2635 (2006)

\bibitem{kv} K.~Vandersloot, Phys. Rev. \textbf{D75}, 023523 (2007)

\bibitem{mb2} M.~Bojowald, Phys. Rev. Lett. \textbf{89} 261301
    (2002)

\bibitem{ps2} P.~Singh, Phys. Rev. \textbf{D73} 063508 (2006)

\bibitem{nott}E.~J.~Copeland, D.~J.~Mulryne, N.~J.~Nunes,
    M.~Shaeri,  Phys. Rev. \textbf{D77}, 023510 (2008);
     Phys. Rev. \textbf{D79}, 023508 (2009)

\bibitem{nott2}Research highlights, Nature Physics 4, 161 (2008)

\bibitem{gre1}J.~Grain, A.~Barrau, Phys. Rev. Lett. \textbf{102},
    081301 (2009)

\bibitem{gre2}J.~Grain, \texttt{arXiv:0911.1625}

\bibitem{gns}C. Germani, W. Nelson and M. Sakellariadou,
    Phys. Rev. \textbf{D76}, 043529 (2007)

\bibitem{jm} J.~Mielczarek, Phys. Rev. \textbf{D81}, 063503 (2010)

\bibitem{mb1} M.~Bojowald, Phys. Rev. Lett. \textbf{86}, 5227-5230
    (2001).

\bibitem{alrev} A.~Ashtekar and J.~Lewandowski, Class. Quant.
    Grav. {\bf 21}, R53-R152 (2004)

\bibitem{crbook} C.~Rovelli,{\em Quantum Gravity}. (Cambridge
    University Press, Cambridge (2004))

\bibitem{ttbook} T.~Thiemann, {\em Introduction to Modern
    Canonical Quantum General Relativity.} (Cambridge University Press,
    Cambridge, (2007))

\bibitem{bp} E.~Bentivegna and T.~Pawlowski, Phys. Rev. D77, 124025
    (2008)

\bibitem{awe2} A.~Ashtekar and E.~Wilson-Ewing, Phys. Rev.
    \textbf{D79}, 083535 (2009)

\bibitem{awe3} A.~Ashtekar and E.~Wilson-Ewing, Phys. Rev.
    \textbf{D}, Phys.Rev. \textbf{D80} 123532 (2009)

\bibitem{we} E.~Wilson-Ewing, Loop quantum cosmology of Bianchi IX
    models (pre-print)

\bibitem{gowdy} M.~Martin-Benito, L.~J.~Garay and G.~A.~Mena
    Marugan, \textit{Phys. Rev.}\textbf{D78}, 083516 (2008)\\
D.~Brizuela, G.~Mena-Marugan and T. Pawlowski, \texttt{arXiv:0902.0697}\\
G.~Mena-Marugan and M.~Martin-Benito, Intl. J. Mod. Phys.
\textbf{A24}, 2820-2838 (2009)

%\bibitem{cs1} A.~Corichi and P.~Singh, Phys. Rev. \textbf{D78},
%    024034 (2008)


%\bibitem{mb-rev} M.~Bojowald, Liv. Rev. Rel. \textbf{8}, 11 (2005).

\bibitem{aa-badhonef} A.~Ashtekar, Gen. Rel. and Grav.
    \textbf{41}, 707-741 (2009)

\bibitem{phenom} A.~Linde, arXiv:hep-th/0503195

\bibitem{ga} D.~Marolf,  \texttt{arXiv:gr-qc/9508015};\\
A.~Ashtekar, J.~Lewandowski, D.~Marolf, J.~Mour\~ao
    and T.~Thiemann, {Jour. Math. Phys.} \textbf{36}, 6456-6493
    (1995)\\
    A.~Ashtekar, L.~Bombelli and A.~Corichi, Phys. Rev. \textbf{D72},
    025008 (2005)

\bibitem{ty} V.~Taveras and N.~Yunes, Phys. Rev. \textbf{D78},
    064070 (2008)




%\bibitem{aps4} A.~Ashtekar, T.~Pawlowski and P.~Singh, Loop quantum
%    cosmology in the pre-inflationary epoch (in preparation)



\end{thebibliography}
\end{document}